\documentclass[twocolumn,prl,showpacs,floatfix]{revtex4}
\usepackage{graphicx}
\usepackage{epsfig}
\usepackage{rotating}
\usepackage{amsmath}
\usepackage{amsfonts}
\usepackage{amssymb}
\usepackage{enumerate}
\usepackage{longtable}
\usepackage{dcolumn}% Align table columns on decimal point
\usepackage{bm}

\begin{document}

\title{Phase Diagram of the $J_1$, $J_2$, $J_3$ 
Heisenberg Models on the Honeycomb Lattice: A Series Expansion Study}

\author{J. Oitmaa }
\affiliation{School of Physics, The University of New South Wales,
Sydney 2052, Australia}

\author{R. R. P. Singh}
\affiliation{University of California Davis, CA 95616, USA}

\date{\rm\today}

\begin{abstract}
We study magnetically ordered phases and their phase boundaries 
in the $J_1-J_2-J_3$ Heisenberg models on the honeycomb lattice
using series expansions
around N\'eel and different colinear and non-colinear magnetic
states. An Ising anisotropy
($\lambda=J_{\perp}/J_z\ne 1$) is introduced and ground state energy 
and magnetization order parameter are calculated as a power seies 
expansion in
$\lambda$. Series extrapolation methods are used to study properties
for the Heisenberg model ($\lambda=1$). We find that at large $J_3$ ($>0.6$)
there is a first order transition between N\'eel and columnar
states, in agreement with the classical answer. For $J_3=0$, we find 
that the N\'eel phase extends beyond the region of classical stability.
We also find that spiral
phases are stabilized over large parameter regions, although their spiral angles can be substantially renormalized with respect to the
classical values. Our study also shows
a magnetically disordered region at intermedaite $J_2/J_1$ and $J_3/J_1$ 
values.

\end{abstract}


\maketitle

\section{Introduction}

The search for quantum spin-liquid phases in realistic spin 
and electronic models and real materials remains a very active
area of research. In recent years a number of experimental materials
have been synthesized, whose behavior has been very promising from
the point of view of discovering such a phase. These include
many kagome and triangular lattice based frustrated magnets.\cite{fukuyama,helton,shimizu,yamashita}
%For example, there is Helium-3 layers absorbed on graphite,\cite{fukuyama}
%the structurally perfect Herbertsmithite kagome systems,\cite{helton} 
%as well as a number of triangular-lattice based
%organic molecular crystals.\cite{shimizu,yamashita}
 In several cases, the frustration parameter,\cite{ramirez}
defined as the ratio of the Curie-Weiss temperature to any magnetic 
ordering temperature exceeds $1000$. Most of these experimental
systems show gapless spin excitations, though the role of impurities 
and weak anisotropies in creating gapless spin excitations has not 
been ruled out.

From a theoretical point of view, frustration is essential for
obtaining a spin-liquid phase in dimensions greater than one. 
Unfrustrated two-dimensional spin models show robust N\'eel order.
However, both spin-frustration arising from competing exchange
interactions and frustration arising from the itineracy of
the electrons can lead to a spin-liquid phase. In recent years
several realistic models have emerged as candidates
for quantum spin-liquid ground states.
One is the Heisenberg model on a kagome
lattice, where extensive Density Matrix Renormalization Group
 work by Yan et al\cite{yan} presents strong
evidence for a spin-liquid phase with a not-too-small gap
to singlet and triplet excitations. Another is the triangular-lattice
antiferromagnet with Heisenberg and 4-spin ring 
exchanges.\cite{motrunich,kpschmidt}
A third, somewhat unexpected case, is the half-filled honeycomb
lattice Hubbard model,\cite{meng} which would lead to an unfrustrated Heisenberg
model at large U. It shows a spin-liquid phase at intermediate
values of U, sandwiched between a N\'eel phase at large U
and a semi-metal phase at small $U$.\cite{paiva} This spin-liquid phase
was also found to be gapped with no signs of any broken symmetry. 

\begin{figure}
\begin{center}
 \includegraphics[width=8cm]{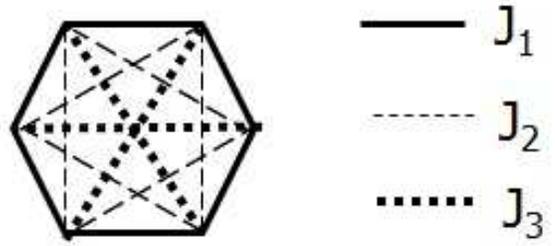}
\caption{\label{fig1} 
Exchange constants of the honeycomb-lattice Heisenberg model
}
\end{center}
\end{figure}

The question of whether this
latter phase can be realized in a frustrated spin
model on the honeycomb lattice, arising from higher order charge
exchange processes has attracted considerable interest.\cite{yang}
Several recent papers have investigated this question using
exact diagonalization, Schwinger-Boson and spin-wave theory,
coupled cluster methods,
and variational wavefunctions. They have come to conflicting
conclusions about the phase diagram and the possible existence
of a spin-liquid phase in this model.\cite{mulder,cabra11,
clark,mosadeq,farnell,reuther,alb}
Experimental realizations of spin-half honeycomb lattice Heisenberg
model have been discussed recently by Tsirlin et al.\cite{tsirlin}

\begin{figure}
\begin{center}
 \includegraphics[width=8cm]{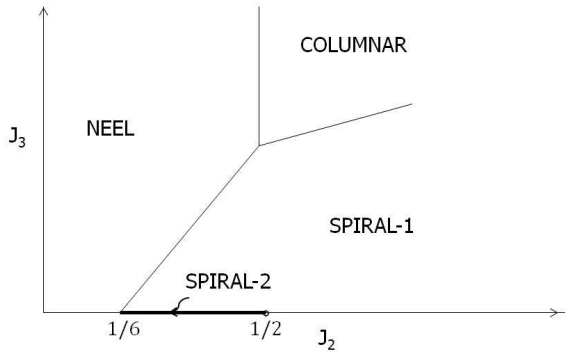}
\caption{\label{fig2} 
Classical phase diagram of the honeycomb lattice Heisenberg
model with $J_1=1$.
}
\end{center}
\end{figure}

Here we study the phase diagram of frustrated Heisenberg
model on the honeycomb lattice using series expansions.\cite{book,advphy}
We will confine our study to all exchanges being positive
($J_1,J_2,J_3>0$) (See Fig.~1). 
The nearest-neighbor honeycomb lattice Heisenberg model,
with antiferromagnetic exchange $J_1$ (which we set to unity),
is well known to have a N\'eel ordered ground state.\cite{oitmaa78,reger,weihong91,oitmaa92,mattsson,fouet} 
Adding
a frustrating second neighbor antiferromagnetic interaction $J_2$ and a third
neighbor antiferromagnetic interaction $J_3$ leads to a complex phase diagram
even at a classical level\cite{rastelli,fouet} 
that includes N\'eel, columnar and different
spiral phases (See Fig.~2). 
We have carried out series expansions around N\'eel,
columnar and spiral phases. For large $J_3$ ($>0.6$) we find
a direct first order transition between N\'eel and columnar phases 
as a function of $J_2$. Thus, the most interesting part of the
phase diagram is for $J_3<0.6$. We have studied this phase
diagram along constant $J_3$ lines as a function of $J_2$ as
well as along the contour $J_3=J_2$. We find that at small $J_3$
the N\'eel phase widens due to quantum fluctuations, in agreement
with recent exact diagonalization study.\cite{alb} We also find that
spiral phases are stabilized for intermediate and large $J_2$.
The spiral angles are strongly renormalized with respect
to their classical values. However, they do not lock into
commensurate colinear structures as suggested in recent exact
diagonalization study.\cite{alb} Rather, our studies favor
angles closer to $90$ degrees, which implies spins pointing at
right angles.

Our study also shows that a part of the phase diagram 
at small and intermediate $J_3$ and $J_2$ has no magnetic 
long range order.
The convergence of our series expansion is best for colinear phases,
when they have clearly non-zero order parameters.
We are unable to accurately obtain the region of stability of
magnetically disordered phases, but it is clear from the general
analysis that such a phase must exist.
This is also clearly seen by
going along $J_2=J_3$, where such a phase exists near the highly
frustrated point $J_2=J_3=0.5$.\cite{cabra11} 
It is possible that a strip of this 
phase extends all the way down to $J_3=0$.

In order to gain some insight into the nature of this phase,
we have also carried out series expansions around a 
dimer state, which was considered as a possible ground state in
Ref.~\onlinecite{mulder}. Following Moessner et al,\cite{moessner} such
a dimer state is called a staggered dimer state. However,
we found that the staggered dimer phase has rather high energy
and thus is not stabilized for $J_3=0$. While our study can not rule
out such a phase for $J_3=0.3$ and intermediate $J_2$, the energies
again suggest that the staggered dimer state has too high an energy.
This should be contrasted with studies of the square-lattice $J_1-J_2$
models, where energies of dimer-phases match smoothly with the
magnetically ordered phases at intermediate $J_2/J_1$ values.\cite{singh}
It is possible that some other dimer or plaquette phase is stable
here.\cite{alb} Investigation of larger unit cell
Dimer/Plaquette phases is left for future work.

\section{Models and Series Expansions}
We study the honeycomb-lattice Heisenberg model with Hamiltonian
\begin{equation}
{\cal H}=J_1 \sum_{<i,j>} \vec S_i \cdot \vec S_j
+J_2\sum_{(i,k)} \vec S_i \cdot \vec S_k
+J_3\sum_{[i,l]} \vec S_i \cdot \vec S_l.
\end{equation}
%where the first sum runs over all nearest-neighbor pairs, the second
%over all second neighbor pairs and the third over all third neighbor pair
%of sites of the honeycomb lattice. 
The exchanges within
an elementary hexagon are shown in Fig.~1. 
Without loss of generality, we take $J_1=1$. 
The classical phase diagram for $J_2>0$, $J_3>0$ is sketched in Fig.~2.
The spin structure of the classical phases are shown in Fig.~3.
It consists of the N\'eel phase, 
a columnar phases, which is also colinear, and two planar
spiral phases. In the spiral-1 phase the pitch vector $\vec Q$ is
perpendicular to one of the bond directions, and is characterized
by a single angle $\theta$. Spins twist away from the N\'eel state
by this angle as one goes from one row to the next. In the 
spiral-II phase the pitch vector is parallel to one of the bond 
directions. It
is characterized by two angles, $\theta$ and $\phi$. Within a unit cell,
the spins deviate from the N\'eel state by an angle $\phi$. In addition,
there is a twist by an angle theta as one goes from one unit cell to
next. The spiral II phase is stabilized by negative $J_3$ but becomes
degenerate with the spiral I phase for $J_3=0$ and $1/6<J_2<1/2$.
Also, at $J_2=1/2$, the spiral-II phase
locks into another colinear phase with $\phi=\pi$, $\theta=\pi$. 

\begin{figure}
\begin{center}
 \includegraphics[width=8cm]{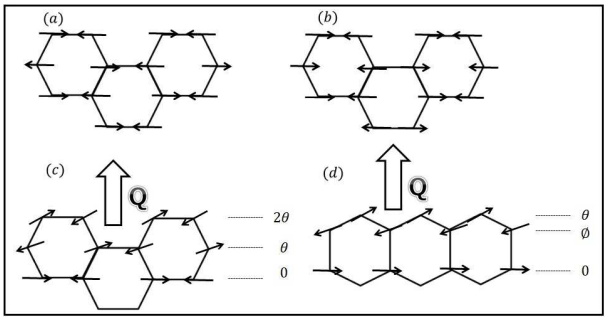}
\caption{\label{fig3} 
Classical phases of the honeycomb-lattice Heisenberg model
with positive $J_1$, $J_2$ and $J_3$.
The phases are (a) N\'eel, (b) columnar, (c) spiral-I
and (d) spiral-II.
}
\end{center}
\end{figure}

To carry out an Ising expansion around a colinear state
(N\'eel or columnar), we introduce
an Ising anisotropy in all the exchange interactions,
\begin{equation}
\vec S_i \cdot \vec S_j = S_i^z S_j^z + \lambda ( S_i^x S_j^x +S_i^y S_j^y).
\end{equation}
We calculate series expansions in the variable $\lambda$
for the ground state energy, per spin, $e_0$ and the sublattice magnetization
$m$.
%defined by 
%\begin{equation}
%{\cal H} |\psi_0> = N e_0 |\psi_0>,
%\end{equation}
%and
%\begin{equation}
%m=<\psi_0| S_i^z |\psi_0>.
%\end{equation}
%We also add a transverse uniform field $-h\sum_i S_i^x$, which allows 
%us to define the uniform susceptibility as
%\begin{equation}
%\chi={\partial^2 e_0\over \partial h^2}.
%\end{equation}

%For the non-colinear states, one needs to rotate the spins, so that the ordering
%vector is pointing along the z-axis at every site and then introduce an
%anisotropy and pick as unperturbed Hamiltonian an Ising term that favors
%this ordered state and treat all other terms as perturbation.\cite{singh-huse} 
%%Such an expansion introduces various additional terms into the
%%Hamiltonian, making the calculations more difficult.

The reader unfamiliar with series expansion methods can find technical details in refs. 25,26 and we shall not elaborate here. However it is worth mentioning that it is convenient, for both colinear and non-colinear states, to rotate the spin axes on different sites to give an unperturbed ground state in which all spins are aligned. This procedure has been described in ref.~\onlinecite{zheng-mckenzie}. 
This transformation has the effect of complicating the form of the Hamiltonian but avoids the need to identify sublattice types in the cluster data.

     To derive series to a given order n requires enumeration of clusters with up to n sites. This number grows rapidly with n, and this is the major factor limiting the length of series. For example, for the 8th order series for the spiral I phase for the full J1-J2-J3 model a total of 1,083,315 distinct clusters occur. For the J1-J2 model an additional term can be obtained, requiring a total of 1,172,204 clusters.

%In all cases, series expansion coefficients for $e_0$, and $m$ were 
%obtained by the linked cluster method. It is well-known that the number
%of graphs (or clusters) increases rapidly with increase in number 
%of bond-types.
%For the N\'eel phase, with $3$ bond types, 67980 distinct clusters 
%were needed to do the complete calculation to $8$-th order.
%For the columnar phase, also with $3$ bond types,
%420,429 distinct clusters were needed to complete
%the calculations to $8$-th order.  For the spiral-1 phase,
%there are $10$ bond-types in the $J_1-J_2-J_3$ model and 1,083,315 
%distinct clusters are needed for a calculation to $8$-th order.

The spiral II phase is only considered for $J_3=0$ (it is not
stabilized classically for positive $J_3$). 
It has $7$ bond-types and took 108,453
distinct clusters to calculate the series to 8th order.
For both the spiral phases,
we allow angles ( such as $\theta$, $\phi$) different from classical 
values and take the angles that minimize the ground state energy. 

In addition, we have also developed series expansion  for the 
ground state energy of the dimer configuration shown in Fig.~4. 
The center of the
dimers form a triangular-lattice with three different types of 
inter-dimer interactions with $J_1$, $J_2$ and $J_3$ all non-zero.
To carry out the calculation to 8-th order
required 33826 linked clusters.

The series data is too lengthy to reproduce here, 
but can be provided on request.

\begin{figure}
\begin{center}
 \includegraphics[width=4cm]{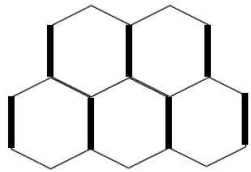}
\caption{\label{fig4} 
Staggered dimer phase on the honeycomb lattice
}
\end{center}
\end{figure}

\section{Series analysis and Results}

We will restrict attention to a 
few contours in the $J_2-J_3$ plane. We will consider $J_3=0$, $J_3=0.3$,
$J_3=0.6$ and $J_3=1$,
and $J_3=J_2$. For large $J_3$ N\'eel and columnar states are
considered, whereas for small $J_3$ N\'eel and spiral phases
are considered.

The series are analyzed by Pad\'e as well as d-log Pad\'e approximants.
These approximants are constructed in the variable $\lambda$ as
well as in the variable $\delta=2\lambda-\lambda^2$. The latter\cite{huse}
allows one to eliminate a square-root singularity due to the gapless
spin-waves at $\lambda=1$.

\begin{figure}
\begin{center}
 \includegraphics[angle=270,width=8cm]{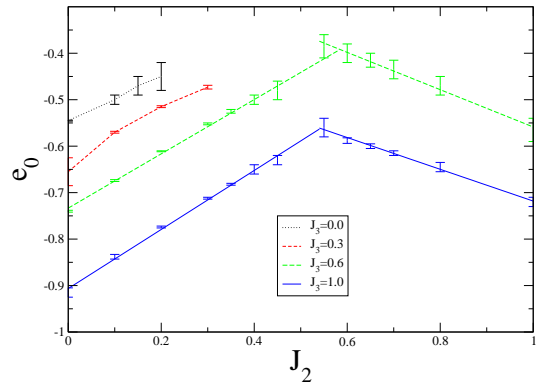}
\caption{\label{fig5} 
Ground state energy estimates of the N\'eel phase ($J_2<0.5$) and
the columnar phase ($J_2>0.5$) are shown 
as a function of $J_2$ for different values of $J_3$.
For $J_3=0.6$ and $J_3=1.0$ linear fits to the data points are also
shown to illustrate the 
first order transition point between the two phases.
}
\end{center}
\end{figure}

\begin{figure}
\begin{center}
 \includegraphics[angle=270,width=8cm]{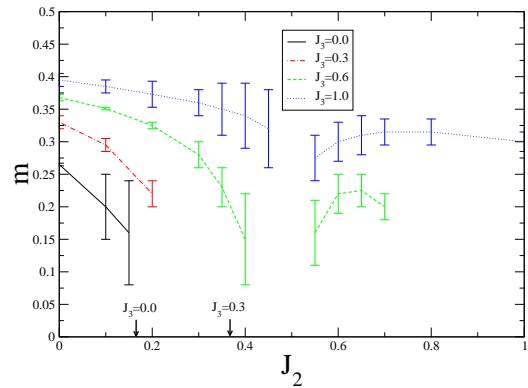}
\caption{\label{fig6} 
Magnetization for N\'eel phase ($J_2<0.5$) and columnar phase ($J_2>0.5$)
as a function of $J_2$ for different values of $J_3$. The arrows 
indicate the critical $J_2$ values for the classical model, where the
N\'eel order parameter vanishes, for $J_3=0.0$ and $J_3=0.3$.
}
\end{center}
\end{figure}

\begin{figure}
\begin{center}
 \includegraphics[angle=270,width=8cm]{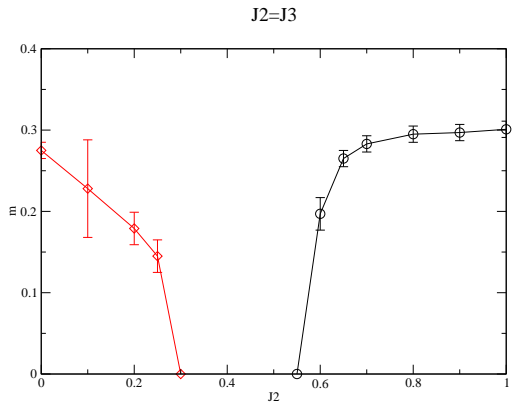}
\caption{\label{fig7} 
Order parameter for the N\'eel and columnar phases along $J_3=J_2$.
}
\end{center}
\end{figure}

In Fig.~5, we show the 
ground state energy of the N\'eel and columnar phases for different
values of $J_3$. The corresponding magnetizations are shown in Fig.~6.
For $J_3=1.0$, there is a clear first order phase transition between the 
N\'eel and columnar phases around $J_2>0.5$. An estimate of the
transition point can be obtained by the linear fits to the energy data
(also shown).
For $J_3=0.6$, the transition from the two sides is 
close to becoming continuous. 
The order parameters for both phases approach zero as $J_2$
approaches $0.5$. Very near the transition, the series analysis
is not reliable enough to tell if this is a continuous
transition, or a first order transition, or there is a small
intermediate region with no magnetic order, although a direct
transition between the phases will necessarily be first order,
as seen by the linear fits to the ground state energy. 

The sharp downturn in the columnar-state magnetization for larger $J_2$
is a clear signature of an impending transition to the spiral I phase,
which classically would occur at $J_2=0.7$.

\begin{table}
  \begin{tabular}{| l | c | c | c| }
  \hline
Phase & $J_2$, $J_3$ &  $\theta_{cl},\phi_{cl}$ & $\theta_r,\phi_r$ \\ \hline \hline
Spiral-I  & 1.00, 0.0 & 104.5   &  90.0   \\ \hline
Spiral-I  & 0.70, 0.0 & 98.2    &  90.0   \\ \hline
Spiral-I  & 0.50, 0.0 & 90.0    &  85.0   \\ \hline
Spiral-I  & 0.70, 0.3 & 104.5   &  95.0   \\ \hline
Spiral-I  & 0.60, 0.3 & 99.6    &  80.0   \\ \hline
Spiral-II & 0.40, 0.0 & 149,125 & 150,95  \\ \hline
Spiral-II & 0.25, 0.0 & 104.5,75.5 & 120,80  \\ \hline
 \end{tabular}
   \caption{\label{spiral-angles} Selected examples of
classical spiral angles ( $\theta_{cl}$ for spiral-I phase
and $\theta_{cl}$ and $\phi_{cl}$ for spiral-II phase)
and the corresponding estimated
renormalized spiral angles ($\theta_r$ and $\phi_r$)
in degrees.
   }
\end{table}

In Fig.~7, we show the magnetization for the N\'eel and columnar
phases along the contour $J_2=J_3$ in the parameter space.
We can see that both order parameters go to zero before the
highly frustrated point $J_2=J_3=0.5$ is reached and there is
an intermediate phase with no magnetic order. This agrees
with previous theoretical study by Cabra et al.\cite{cabra11}

\begin{figure}
\begin{center}
 \includegraphics[angle=270,width=8cm]{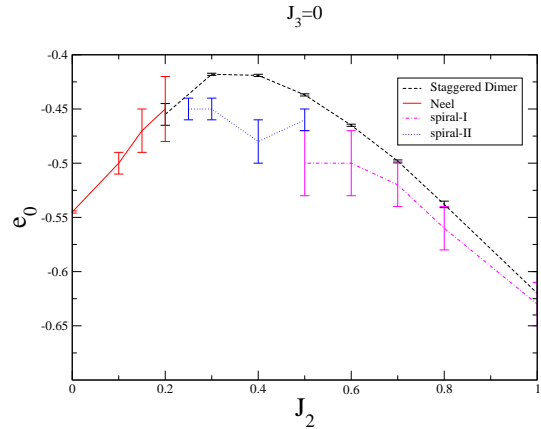}
\caption{\label{fig8} 
Ground state energy for N\'eel, staggered  dimer, spiral-I and 
spiral-II phases for $J_3=0$.
}
\end{center}
\end{figure}

\begin{figure}
\begin{center}
 \includegraphics[angle=270,width=8cm]{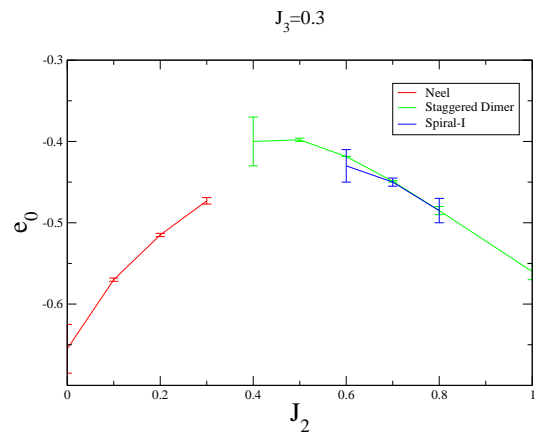}
\caption{\label{fig9} 
Ground state energy for N\'eel, staggered dimer and spiral-I phases
for $J_3=0.3$.
}
\end{center}
\end{figure}

At small $J_3$ (for example $J_3=0$)
the region of N\'eel order is increased with respect to the classical 
answer, but that is no longer true for $J_3=0.3$. For $J_3=0$,
we estimate that N\'eel order extends upto $J_2=0.20$ compared
to the classical value of $1/6$. In contrast,
for $J_3=0.3$, the magnetization
vanishes near $J_2=0.3$ compared to the classical value of
$11/30$ (See Fig.~6). We do not find any clear
evidence for a magnetically disordered phase at $J_3=0$. Fig.~8 shows
the ground state energy for N\'eel, spiral-I, spiral-II and
staggered dimer phases for $J_3=0$. In the spiral phases, we consider a
range of spiral angles and pick that value which minimizes the energy.
In general, these energy functions are quite shallow, so that the
renormalized spiral angles shown in Table-I should be considred as
approximate. While the convergence is not
excellent at intermediate $J_2/J_1$ values, our analysis 
suggests that one transitions from N\'eel to spiral-II
to spiral-I phase. The staggered dimer phase is clearly not
stabilized here in contrast to recent suggestions
from bond-operator formalism\cite{mulder} based studies.
In addition, the spiral-II phase also does not lock 
into a colinear configuration in contrast to the conclusions
from the finite-size study.\cite{alb}
If anything, our estimates
suggest spiral angles close to $90$ degrees, which implies some spins pointing
at right angles with respect to their neighbors.

In Fig.~9, we show the ground 
state energy for N\'eel, spiral-I, and staggered dimer
configurations along $J_3=0.3$, as a function
of $J_2$. At intermediate $J_2$ values the convergence
of both N\'eel and spiral-I phases becomes poor. The
energy of the staggered dimer phase appears well behaved.
In this case, there is a much stronger case for an
intermediate phase with no magnetic order.
However, the staggered dimer phase
has a relatively high energy and it is not clear that
it becomes the ground state of the system in this region.
Since the ground state energy must vary continuously, it
seems more likely that some other phase not considered by us
becomes the ground state in this region. At large $J_2$,
the energy of spiral and staggered dimer phases become
nearly degenerate. Our results are consistent with the suggestion
from the exact diagonalization study\cite{alb} that for some parameter
ranges, the staggered dimer state may be difficult to distinguish
from magnetically ordered states.

\section{Discussions and Conclusions}

In this paper, we have used series expansion methods to study the ground
state phase diagram of the $J_1-J_2-J_3$ antiferromagnetic Heisenberg models on the honeycomb lattice. We have determind the stability of N\'eel,
columnar and various spiral phases and calculated their properties.
In agreement with previous studies, we find that the region of
stability of N\'eel phase increases with quantum fluctuations
for $J_3=0$. However, that is not the case for larger $J_3$.
There clearly exists a parameter region at intermediate
$J_2$ and $J_3$ with no magnetic order. This region
is most clearly seen near the highly frustrated point $J_3=J_2=0.5$
and possibly forms a strip extending all the way near to $J_3=0$.
We found that the staggered dimer order is not favored for the $J_1-J_2$
model. 

The phase diagram has some resemblence to the square-lattice Heisenberg
model with frustrated antiferromagnetic exchange constants. 
In all these
systems, it remains difficult to determine the nature of the magnetically 
disordered phases. 
Our results are generally in good agreement with recent
exact diagonalization study of Albuquerque et al.\cite{alb}
One point of difference is that we do not find 
that a third colinear phase (that is the spiral II with $\phi=\pi$,
$\theta=\pi$) is favored over general spiral phases over any extended
part of the phase diagram. This difference may be because periodic
boundary conditions on finite systems may disfavor incommensurate
phases. Another difference with respect to the Bond Order
Mean-Field Theory study \cite{mulder} is that we do not find evidence for
a staggered dimer phase in the magnetically disordered region at small $J_3$.

An important question is whether this magnetically disordered 
phase in the frustrated Heisenberg models is related to the one found
by Meng et al\cite{meng} in the Hubbard model at intermediate $U/t$,
and whether it is a true spin-liquid. 
Recent work by Yang and Schmidt\cite{yang}
found that for $U/t$ values where Meng et al obtained a phase transition
to a spin-liquid phase $J_2/J_1$ remains very small (of order $0.06$),
with $J_3/J_1$ even smaller. This means that just these additional
interactions can not drive the transition, as our study finds that
these parameters are well within the N\'eel phase. 
It is possible that larger ring
exchanges play a role in bringing about the transition. 
Nevertheless, the disordered phase in the
Heisenberg model may be connected to the one found in the Hubbard
model. This question as well as the possibility of other types of dimer/plaquette phases deserves further attention.

\begin{acknowledgements}
This work is supported in part by NSF grant number  DMR-1004231.
\end{acknowledgements}

%\bibliographystyle{apsrev}
%\bibliography{../bibinput/liter10}

\end{document}